# Beyond Bounds on Light Scattering with Complex Frequency Excitations

Seunghwi Kim[1], Sergey Lepeshov[2], Alex Krasnok[3], and Andrea Alù[1,4*]

[1]Photonics Initiative, Advanced Science Research Center, City University of New York, New York, New York 10031, USA

[2]DTU Electro, Department of Electrical and Photonics Engineering, Technical University of Denmark, Ørsteds Plads 343, Kgs. Lyngby, DK-2800, Denmark

[3]Department of Electrical and Computer Engineering, Florida International University, Miami, FL 33174, USA

[4]Physics Program, Graduate Center, City University of New York, New York, NY 10016, USA

*Corresponding author: aalu@gc.cuny.edu

*Light scattering is one of the most established wave phenomena in optics, lying at the heart of light-matter interactions and of crucial importance for nanophotonic applications. Passivity, causality and energy conservation imply strict bounds on the degree of control over scattering from small particles, with implications on the performance of many optical devices. Here, we demonstrate that these bounds can be surpassed by considering excitations at complex frequencies, yielding extreme scattering responses as tailored nanoparticles reach a quasi-steady-state regime. These mechanisms can be used to engineer light scattering of nanostructures beyond conventional limits for noninvasive sensing, imaging, and nanoscale light manipulation.*

The study of light scattering by small particles has a long history [1], and it is the basis of a disparate range of phenomena, from the color of the sky to the brightness of stained glasses. Despite its well-established nature, engineering light scattering is not an outdated problem: it remains of paramount interest in photonics [2,3], not only enabling new discoveries [4,5], but also broadening the impact of nanophotonics for various applications, such as optical antennas [6], imaging [7], optical tweezers and trapping [8,9].

Small nanoparticles are typically characterized by a broad donut-shape scattering pattern sustained by their dominant dipolar fields. More exotic scattering features can be achieved in suitably tailored nanoparticle geometries by carefully balancing electric and magnetic dipolar scattering, yielding destructive interference in specific directions that make the scattering more directive. For instance, scattering suppression in either the forward or backward direction can be obtained in nanoparticles satisfying the Kerker conditions [10]. Magneto-electric, plasmonic [11] and high-index dielectric (DNPs) [12,13] nanoparticles have been shown to support peculiar scattering features, offering opportunities for light manipulation, and establishing the basis for the design of metasurfaces and metamaterials. However, these exotic responses require careful design of the nanoparticle geometries, and they emerge only at specific wavelengths as a function of the available material dispersion. In addition, even when these conditions are met, the scattering from nanoparticles remains limited by causality, passivity and energy conservation.

Consider, for instance, the problem of realizing a particle with directional backward scattering, i.e., with minimized forward scattering cross-section $\sigma_F = \frac{\pi}{|k_i|^2}\left|\sum_{n=1}^{\infty}(2n+1)(a_n+b_n)\right|^2 \simeq 0$, where $k_i = |\mathbf{k}_i|$ is the wavenumber of the incident wave, $n$ is the multipolar order, and $a_n$ and $b_n$ are electric and magnetic Mie scattering coefficients, corresponding to the amplitude of multipolar scattered waves of order $n$ [2,14]. We assume an $e^{-i\omega t}$ time convention under plane wave excitation $\mathbf{E}_{in} = E_o e^{-i\mathbf{k}_i \cdot \mathbf{z}}\hat{\mathbf{x}}$, where $\hat{\mathbf{x}}$ is a unit vector. According to the optical theorem, the extinction cross-section $\sigma_{ext}$ is proportional to the normalized forward scattering amplitude with polarization parallel to the incident wave $\hat{\mathbf{e}}_i \cdot \mathbf{f}(\mathbf{k}=\mathbf{k}_i)$ [2,15]:

$$\sigma_{ext} = \sigma_{abs} + \sigma_{scat} = \frac{4\pi}{k_i} \text{Im}\left[\hat{\mathbf{e}}_i \cdot \mathbf{f}(\mathbf{k}=\mathbf{k}_i)\right], \tag{1}$$

where $\sigma_{abs}$ and $\sigma_{scat}$ are the absorption and scattering cross-sections, $\hat{\mathbf{e}}_i$ is the unit polarization vector of the incident wave, and $\mathbf{k}_i$ and $\mathbf{k}$ are the wave vectors of incident and scattered waves, respectively. Eq. (1) indicates that $\sigma_{ext}$ must be zero if $\mathbf{f}(\mathbf{k}=\mathbf{k}_i)=0$, yet in a passive scatterer $\sigma_{abs} \geq 0$, $\sigma_{scat} \geq 0$, hence this condition can be met only if the scattered power is zero at all angles [15-18]. For very small particles, a negligible – yet strictly nonzero – forward scattering can be achieved, with most of the residual scattering distributed at other angles, as in the inset of Fig. 1(a), but a severe trade-off exists between total scattering and residual forward scattering, consistent with Eq. (1) [16]. Passivity fundamentally limits how backward directive a scattering pattern can be.

Active materials can be employed to overcome this limitation, since such particles can support $\sigma_{abs} < 0$, relaxing the constraint on zero forward scattering. Gain provides additional energy, relaxing limitations that stem from power conservation, and offering a broader control over light scattering. However, it is challenging to embed active media within nanophotonic systems, and typically optical gain comes at the price of stringent bandwidth and stability limitations [19,20]. Moreover, active materials are characterized by unavoidable noise in the form of amplified spontaneous emission or parasitic harmonics in the case of parametric gain [20], which hinder the practical feasibility of active systems for several applications.

The limitations outlined so far implicitly refer to monochromatic excitations. We have recently explored the exotic response of passive nanophotonic systems when excited with signals that oscillate at complex frequencies. Under suitable conditions, we have shown that a tailored resonant linear system excited by a signal oscillating at a complex $\omega$ can reach a *quasi-steady state* response, such that its output after a transient oscillates at the same complex frequency as the input. In this regime, the passivity constraints associated with real frequencies can be overcome, opening new frontiers for light-matter interactions [21]. For instance, based on these principles, we have recently demonstrated a phenomenon analogous to absorption emerging in lossless structures excited at complex frequencies [22,23], with opportunities for efficient energy storage and optical memories [24]. Analogously, coherent excitations at complex frequencies can realize pulling forces, mimicking the emergence of gain [25], which can also be exploited to realize parity-time symmetric phenomena in passive systems [26].

Here we explore the opportunities that complex frequency excitations open to overcome long-held bounds on scattering. First, we demonstrate that a tailored dielectric sphere can support identically zero forward scattering in its quasi-steady state when excited with a signal oscillating at a suitable complex frequency. Figure 1(a) shows the position of the first zero of the forward scattering efficiency, $Q_F = \sigma_F / \pi a^2$, where $a$ is the sphere radius, evaluated in the complex frequency plane for passive DNPs with different refractive indexes. The zero always lies in the lower complex half-plane, and it can emerge close to the real frequency axis for large refractive indexes. For instance, when $m = 6$ (blue circle in the figure), excitation at the real part of this complex frequency produces the scattering pattern shown in the inset, which is asymmetric with minimized forward scattering and points towards the backward direction. The zero moves farther away from the real axis for lower indexes, implying that a monochromatic excitation at the real part of its frequency generates a larger and larger relative forward scattering.

The forward scattering zeros may be pushed towards the real axis by adding gain to the particle material. The resulting signal amplification corresponds to negative absorption, ensuring that the extinction $\sigma_{ext}$ can be made identically zero in Eq. (1) even when $\sigma_{scat} > 0$. In [27] (Fig. S3a)), for instance, we show that for the complex refractive index $m = 3.950 - i0.405$, the first forward scattering zero lies on the real axis. Figure 1(b) compares the normalized scattering patterns in the two polarization planes at the frequency of this forward zero $\omega_{gain}a/c = 0.831$ for the active particle (dotted lines), and compares it to the passive scenario $m = 3.929$ [31], evaluated at $\omega_{nogain}a/c = 0.853$, which corresponds to the real part of the complex frequency zero. We plot half of the angular spectrum for each scenario since the patterns are symmetric, and the red (blue) line indicates the scattered intensity parallel (perpendicular) to the scattering plane. The scattered fields in the forward direction are entirely suppressed in the active particle, while residual fields are still present in the forward scattering for the passive particle.

Interestingly, we can engage the forward scattering zero of a passive particle without relying on material gain, but by exciting it with a signal oscillating at the proper complex frequency. Fig. 2(a) shows the forward scattering efficiency in the complex frequency plane for a sphere with refractive index $m = 3.929$. We can observe a forward scattering pole (bright area) and a zero (dark area) in the plotted range of frequencies. We excite the particle with an incident signal with electric field oscillating at a complex frequency $E_i(t) = E_o \exp(-i\operatorname{Re}[\omega]t)\exp(\operatorname{Im}[\omega]t)$, characterized by an exponentially growing or decaying envelope as a function of the sign of $\operatorname{Im}[\omega]$. Here, we excite the sphere with a decaying signal oscillating at $\omega a/c = 0.824 - i\,0.063$, corresponding to the complex frequency of the forward scattering zero marked by the blue circle. Excitations at complex frequencies are unbounded at $\pm\infty$, and hence they are limited to finite temporal intervals. While no one prevents us from analytically continue the expression of the forward scattering efficiency in the complex frequency plane, as we do in Fig. 2(a), we cannot generally expect that the sphere after a transient reaches a steady-state response oscillating at the same (complex) frequency as the excitation. However, as we show in the following, for excitation of zeros and poles sufficiently close to the real axis, the response of the structure after a short transient can support a quasi-steady state regime in which the scattered fields oscillate at the same complex frequency. Under this condition, we can find exotic scattering responses as predicted by the singularities in the complex frequency plane.

We performed full-wave time-domain simulations to obtain the temporal evolution of the scattered fields, assuming an excitation at $\omega a/c = 0.824 - i\,0.063$ starting at $t = 0$. After a transient that depends on the way we excited the system for $t < 0$, we observe that the forward scattering converges to zero and the scattering from the particle reaches a quasi-steady state response (see time evolution in [27]). The normalized scattering pattern in this quasi-steady-state regime retrieved from FDTD simulations (solid line) is compared to the analytically calculated pattern (dashed line) in Fig. 2(b), confirming that the scattering is directed backwards, with minimum forward scattering orders of magnitude smaller than the minimum allowed for monochromatic excitations [27]. We stress that this result is achieved with a passive particle, and this exotic response can be explained based on the concept of 'virtual gain', enabled when suitably tailored resonances are engaged with exponentially decaying signals oscillating at tailored complex frequencies [25,26]. The sphere scatters in time the energy stored from earlier cycles, for which the input signal was stronger due to its decaying nature. At this complex frequency, the electric and magnetic dipoles reach quasi-stationary states oscillating out of phase with respect to each other, i.e., $a_1 = -b_1$ – a condition that cannot be achieved in a passive scatterer under monochromatic excitation – but which can be obtained here because the dipoles are radiating energy stored at earlier times. In the quasi-steady state, they perfectly cancel each other in the forward direction while interfering constructively in the backward direction.

In order to visualize the phenomenon, we plot the relation between normalized forward scattering cross-section $\sigma_F^{norm} = \frac{\pi \sigma_F}{9 \lambda_o^2}$ and normalized total scattering cross-section $\sigma_{scat}^{norm} = \frac{\pi \sigma_{scat}}{3 \lambda_o^2}$, where $\lambda_o$ is the incident wavelength, in Fig. 2(c). The allowed forward scattering cross-section in a small particle is bounded by the shaded region in the figure, i.e., it cannot be too small without minimizing the total scattering (see further analysis in [27]). The minimum scattering cross-section for different refractive indexes $m = 8, 6, 5$, and $3.93$ are located on the curve under harmonic plane wave illumination. Here, we largely surpass this bound using a complex frequency excitation: the green line in Fig. 2(c) shows the evolution of the scattering response in this plane as we sweep the frequency from $\omega a/c = 0.853$ (on the right side of the figure) to $\omega a/c = 0.824 - i\,0.063$ (on the left side of the figure) by linearly interpolating the real and imaginary parts. The green curve representing the evolution of the forward scattering converges to the *y*-axis, enabling a large scattering cross-section and zero forward scattering in the quasi-steady state.

Complex frequency excitations can also enable other exotic scattering responses. For instance, by tailoring at the same time dipolar and quadrupolar scattering harmonics in the complex frequency plane, we can induce the cancellation of scattered fields in *both* forward and backward directions, enabling transverse scattering patterns [32,33]. Transverse scattering from DNPs has been explored in [34], but without complete suppression of forward and backward scattering because of the mentioned passivity limitations (see Fig. S7 in [27]). This response arises because the sphere is near an anapole resonance due to the destructive interference of electric dipole and quadrupole responses. The overall response makes the total scattered power very small [34], consistent with the requirement of minimizing the forward scattering in Eq. (1). Fig. 3(a) plots the forward scattering efficiency $Q_F$ in the complex plane around the frequency where transverse scattering happens, i.e., near $\omega a/c = 1.1$ and refractive index $m = 3.929$, indeed finding a forward scattering zero in the lower complex half-plane at $\omega a/c = 1.101 - i0.014$. Because of the close interactions of electric dipole and quadrupole harmonics around this resonance, the backward scattering efficiency $Q_B = \sigma_B/\pi a^2$, where $\sigma_B$ is the backward scattering cross-section, also supports a complex zero at a close location in the complex frequency plane, as shown in Fig. 3(a). By exciting the sphere with $\omega a/c = 1.101 - i0.014$, we indeed obtain the scattering patterns shown in Fig. 3(b), where we compare analytical calculations and FDTD simulations again. The results confirm zero forward and backward scattering and a purely transverse scattering pattern in one plane of polarization.

Another unique scattering constraint that stems from passivity is the maximum scattered power associated with a single scattering harmonic. Consider, for instance, the scattering from a dielectric cylinder under transverse-magnetic (TM) plane wave illumination at frequency $\omega_{in}$. The contribution to the scattering width from the $n^{\text{th}}$ multipolar order $\left(\sigma_{scat}^{TM}\right)_n = \dfrac{2\lambda}{\pi}|a_n|^2$, where $a_n$ is the TM Mie scattering coefficient for dielectric cylinders [2]. The maximum contribution to the width for a single harmonic is limited to $2\lambda/\pi$ for passive scatterers in the case of monochromatic excitation since $|a_n| \leq 1$. For complex frequency excitation, however, this bound can be largely surpassed. Specifically, $k$ is complex in the quasi-steady state, so that it can make the Mie coeffcients $a_n$ much larger than unity (see further discussion in [27]). By targeting a pole of the total scattering cross-section in the complex frequency plane, it is possible to realize lasing-like behaviors, exciting a *scattering pole*, as the decay rate of the incoming signal, tailored to match the eigenmode resonance of the cylinder, compensates for the radiation loss that bounds the scattering coefficient for monochromatic excitations.

For example, we consider a cylindrical DNP with $m = 3.929$ and evaluate the poles (yellow circles) of $Q_{scat} = \sum_{n=-\infty}^{\infty} \left(\sigma_{scat}^{TM}\right)_n \bigg/ 2a$ shown in Fig. 4(a), where $a$ is the radius of the cylindrical particle [27]. The third pole, marked by a hollow blue circle at $\omega a/c = 1.250 - i0.0175$, corresponds to a magnetic quadrupolar mode, whose resonance emerges for monochromatic excitations around the real frequency $\omega a/c = 1.248$. The scattering intensity of the cylindrical DNP $I_{scat}(r_o, \theta, t)$ at distance $r_o$ reads

$$I_{scat}(r_o, \theta, t) = I_{in}(r_o, t) \left| \sum_{n=-\infty}^{\infty} (-i)^n a_n H_n^{(1)}(kr_o) \exp(in\theta) \right|^2. \qquad (2)$$

Here, $\theta$ is the polar angle, $k$ is the incident wavenumber, and $H_n^{(1)}(z)$ are the Hankel functions of the first kind. Note that $I_{in}(r_o, t)$ is the input intensity, including an exponentially decaying contribution. We define the normalized scattering pattern $S_{scat}(r_o, \theta, t) = I_{scat}(r_o, \theta, t)/I_{in}(r_o, t)$ as the ratio of scattering intensity $I_{scat}(r_o, \theta, t)$ to the input intensity $I_{in}(t)$ evaluated at the same instant in time. In the quasi-steady state, this function is independent of time, since the scattered fields oscillate at the same complex frequency of the input, thus

$$S_{scat}(r_o,\theta) = \left| \sum_{n=-\infty}^{\infty} (-1)^n a_n H_n^{(1)}(kr_o) \exp(in\theta) \right|^2.$$

We verify with FDTD simulations that we reach a quasi-steady state response exciting at the complex frequency of this scattering pole, as discussed in [27]. Figure 4(b) shows the far-field scattering patterns normalized to the input intensity as time evolves. The scattered intensity decays exponentially in time following the complex frequency excitation, but the ratio $S_{scat}(r_o,\theta,t)$ increases over time, yielding in the quasi-steady state values much larger than 1, which is the bound for monochromatic excitations. At this complex frequency, we reach a purely quadrupolar pattern with scattered fields significantly larger than the input fields at the same instant in time, well beyond the passivity bound. The scattering efficiency at $\tau = t\,\text{Re}[\omega]/(2\pi) = 75.4$ is $Q_{scat} = \sum_{n=-\infty}^{\infty} \frac{\lambda}{\pi a} |a_n|^2 = 520.9$, over two orders of magnitude larger than its value at the quadrupolar resonance for real frequency excitation, $Q_{scat} = 3.6$ in Fig. S9a of [27].

Overall, our results demonstrate that complex frequency excitations can manipulate the scattering of passive resonant objects in exotic ways, going well beyond the limits imposed by passivity. This principle may readily be translated to more complex systems, such as lattices, particle clusters, and metamaterials, and it can be extended to other wave domains, e.g., acoustic systems. Our results demonstrate that in non-monochromatic settings in which we can tailor the input signal profile in time, the response of a scatterer is not limited by passivity, causality, and energy conservation. These phenomena can be applied to various settings, such as noninvasive sensors, energy-storage, efficient wireless power transfer, directional light source, beam control, high-performance antennas, imaging and more. Our works can also be expanded to be combined with PT-symmetry and exceptional point physics [26], yielding high sensitivity and exotic scattering features.

This work was supported in part by the Simons Foundation and the Air Force Office of Scientific Research.

**Figures**

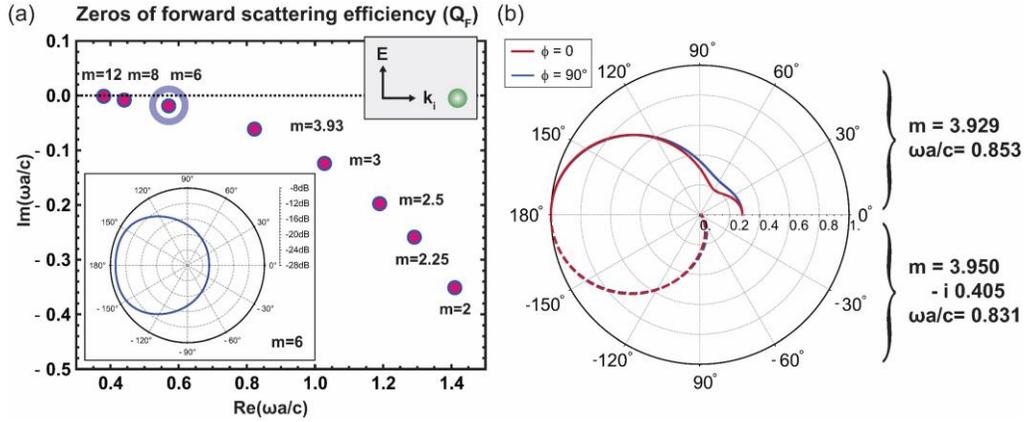

**Figure 1** (a) Evolution of the first zero of the forward scattering efficiency $Q_F$ as we vary the refractive index $m$ of a dielectric sphere. The scattering pattern for $m = 6$ (blue circle) is shown in the inset for excitation at the real part of the complex frequency of the zero. (b) Normalized scattering patterns in the two polarization planes for $m = 3.929$ (solid lines) and $m = 3.950 - i\,0.405$ (dotted lines). Red (blue) lines indicate that the incident waves are parallel (orthogonal) to the scattering plane.

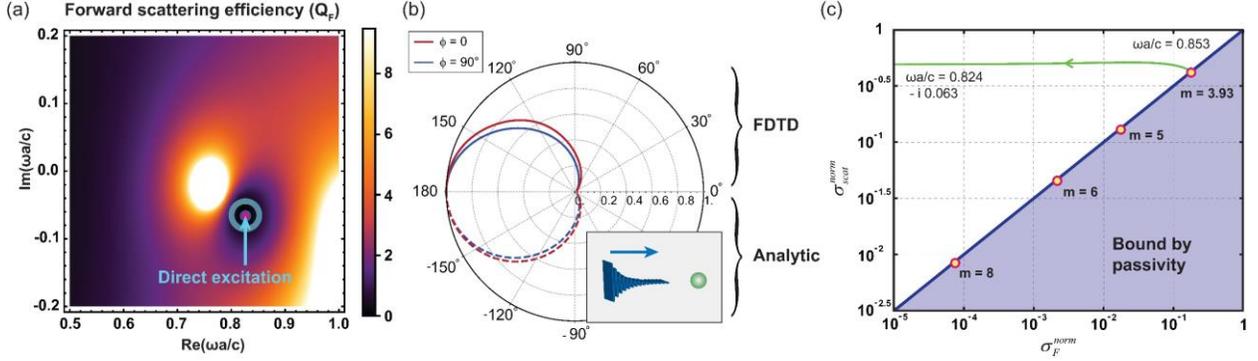

**Figure 2** (a) Density plot of the forward scattering efficiency $Q_F$ in the complex frequency plane. The zero (darkest spot) and pole (brightest spot) are shown in the plot, and the zero is of interest here, marked by the blue circle. (b) Scattering pattern at the complex frequency $\omega a/c = 0.824 - i\,0.063$, demonstrating zero forward scattering in a passive dielectric sphere. The complex excitation impinging on a subwavelength dielectric sphere is schematically illustrated in the inset. These results are retrieved from FDTD simulations (solid lines) and validate our analytical results (dashed lines). (c) The normalized forward and total scattering cross-sections are bounded by the thick blue line for any passive DNPs. The cross-sections for minimizing forward scattering at m = 8, 6, 5 and 3.93 are placed on the bound (the thick blue line). Exciting in the complex frequency plane, as we evolve $\omega a/c = 0.853$ from to $\omega a/c = 0.824 - i\,0.063$, allows us to go beyond the bound (the green line).

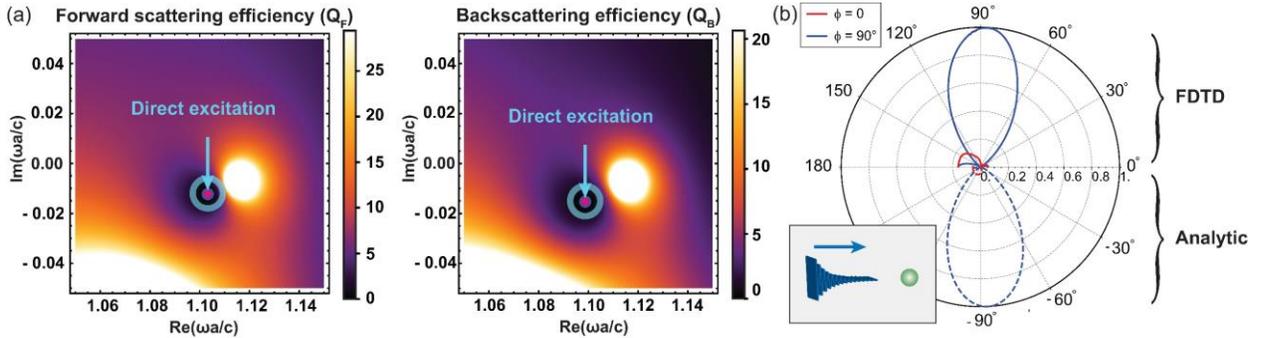

**Figure 3** (a) Density plot of the scattering efficiencies in the forward and backward directions to determine the complex zeros of $Q_F$ and $Q_B$ (marked by the blue circle) in the complex plane. (b) Scattering patterns obtained via FDTD and analytical solutions at the complex frequency ensure zero forward and backward scattering. The simulation and analytical results are in agreement in the quasi-steady state.

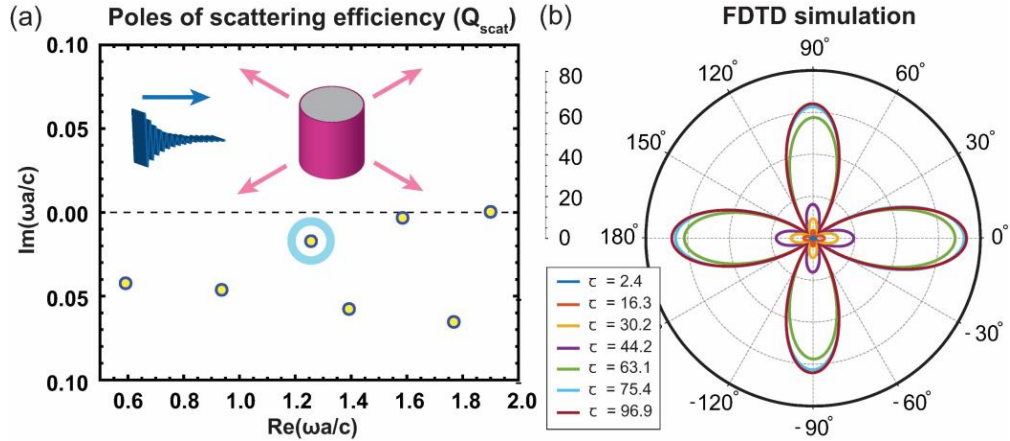

**Figure 4** (a) Poles (yellow circles) of $Q_{scat}$ of each mode are illustrated. The third pole marked by the blue circle is of interest for investigating the scattering pole. (b) Temporal evolution for complex frequency excitations measured at $\tau = t\,\omega/(2\pi) =$ 2.4, 16.3, 30.2, 44.2, 63.1, 75.4, and 96.9 from FDTD simulations.

## References

[1] G. Mie, Beiträge zur Optik trüber Medien, speziell kolloidaler Metallösungen, Ann. Phys. **330**, 377 (1908).
[2] C. F. Bohren and D. R. Huffman, *Absorption and Scattering of Light by Small Particles* (John Wiley & Sons, 2008).
[3] I. L. Fabelinskii, *Molecular scattering of light* (Springer Science & Business Media, 2012).
[4] M. I. Tribelsky and B. S. Luk'yanchuk, Anomalous Light Scattering by Small Particles, Phys. Rev. Lett. **97**, 263902 (2006).
[5] X. Fan, W. Zheng, and D. J. Singh, Light scattering and surface plasmons on small spherical particles, Light Sci. Appl. **3**, e179 (2014).
[6] P. Mühlschlegel, H.-J. Eisler, O. J. F. Martin, B. Hecht, and D. W. Pohl, Resonant Optical Antennas, Science **308**, 1607 (2005).
[7] J. Y. Lee *et al.*, Near-field focusing and magnification through self-assembled nanoscale spherical lenses, Nature **460**, 498 (2009).
[8] A. Ashkin, J. M. Dziedzic, J. E. Bjorkholm, and S. Chu, Observation of a single-beam gradient force optical trap for dielectric particles, Opt. Lett. **11**, 288 (1986).
[9] S. Lal, S. E. Clare, and N. J. Halas, Nanoshell-Enabled Photothermal Cancer Therapy: Impending Clinical Impact, Acc. Chem. Res. **41**, 1842 (2008).
[10] M. Kerker, D. S. Wang, and C. L. Giles, Electromagnetic scattering by magnetic spheres, J. Opt. Soc. Am. **73**, 765 (1983).
[11] B. Luk'yanchuk, N. I. Zheludev, S. A. Maier, N. J. Halas, P. Nordlander, H. Giessen, and C. T. Chong, The Fano resonance in plasmonic nanostructures and metamaterials, Nat. Mater. **9**, 707 (2010).
[12] M. Nieto-Vesperinas, R. Gomez-Medina, and J. J. Saenz, Angle-suppressed scattering and optical forces on submicrometer dielectric particles, J. Opt. Soc. Am. A **28**, 54 (2011).

[13] Y. H. Fu, A. I. Kuznetsov, A. E. Miroshnichenko, Y. F. Yu, and B. Luk'yanchuk, Directional visible light scattering by silicon nanoparticles, Nat. Commun. **4**, 1527 (2013).
[14] H. C. van de Hulst, *Light Scattering by Small Particles* (Courier Corporation, 1981).
[15] J. D. Jackson, (American Association of Physics Teachers, 1999).
[16] A. Alu and N. Engheta, How does zero forward-scattering in magnetodielectric nanoparticles comply with the optical theorem? , J. Nanophotonics **4**, 041590 (2010).
[17] R. Fleury, J. Soric, and A. Alù, Physical bounds on absorption and scattering for cloaked sensors, Phys. Rev. B **89**, 045122 (2014).
[18] R. Fleury, F. Monticone, and A. Alù, Invisibility and Cloaking: Origins, Present, and Future Perspectives, Phys. Rev. Appl. **4**, 037001 (2015).
[19] B. Nistad and J. Skaar, Causality and electromagnetic properties of active media, Phys. Rev. E **78**, 036603 (2008).
[20] A. Krasnok and A. Alù, Active Nanophotonics, Proc. IEEE **108**, 628 (2020).
[21] A. Krasnok, D. Baranov, H. Li, M.-A. Miri, F. Monticone, and A. Alú, Anomalies in light scattering, Adv. Opt. Photonics **11**, 892 (2019).
[22] D. G. Baranov, A. Krasnok, and A. Alù, Coherent virtual absorption based on complex zero excitation for ideal light capturing, Optica **4**, 1457 (2017).
[23] G. Trainiti, Y. Ra'di, M. Ruzzene, and A. Alù, Coherent virtual absorption of elastodynamic waves, Sci. Adv. **5**, eaaw3255 (2019).
[24] M. Cotrufo and A. Alù, Excitation of single-photon embedded eigenstates in coupled cavity-atom systems, Optica **6**, 799 (2019).
[25] S. Lepeshov and A. Krasnok, Virtual optical pulling force, Optica **7**, 1024 (2020).
[26] H. Li, A. Mekawy, A. Krasnok, and A. Alù, Virtual Parity-Time Symmetry, Phys. Rev. Lett. **124**, 193901 (2020).
[27] See Supplementary Material at [URL] for details on the overview of the anti-Kerker effects in active systems, the analysis of the virtual anti-Kerker effects, the virtual generalized Kerker efffects and the scattering pole, and the description of the numerical method. Addditional discussions on the quasi-steady state responses, the passivity bound, and the introduction to the scattering parameters, which include Refs. [28-30].
[28] A. Ishimaru, Wave propagation and scattering in random media (Academic press New York, 1978), Vol. 2.
[29] C. Mätzler, MATLAB functions for Mie scattering and absorption, version 2, (2002).
[30] M. Kerker, The scattering of light and other electromagnetic radiation: physical chemistry: a series of monographs (Academic press, 2013), Vol. 16.
[31] It is the refractive index of silicon at λ ~ 600 nm, which is a possible operating wavelength for experimental realization.
[32] Y. X. Ni, L. Gao, A. E. Miroshnichenko, and C. W. Qiu, Controlling light scattering and polarization by spherical particles with radial anisotropy, Opt. Express 21, **8**091 (2013).
[33] J. Y. Lee, A. E. Miroshnichenko, and R.-K. Lee, Simultaneously nearly zero forward and nearly zero backward scattering objects, Opt. Express 26, **3**0393 (2018).
[34] H. K. Shamkhi *et al., T*ransverse Scattering and Generalized Kerker Effects in All-Dielectric Mie-Resonant Metaoptics, Phys. Rev. Lett. 122**, 1**93905 (2019).